# Is It Possible to Simultaneously Achieve Zero Handover Failure Rate and Ping-Pong Rate?

Hyun-Seo Park, Yong-Seouk Choi, Tae-Joong Kim, Byung-Chul Kim, and Jae-Yong Lee

*Abstract*—Network densification through the deployment of large number of small cells has been considered as the dominant driver for wireless evolution into 5G. However, it has increased the complexity of mobility management, and operators have been facing the technical challenges in handover (HO) parameter optimization. The trade-off between the HO failure (HOF) rate and the ping-pong (PP) rate has further complicated the challenges. In this article, we proposed "ZEro handover failure with Unforced and automatic time-to-execute Scaling" (ZEUS) HO. ZEUS HO assures HO signaling when a user equipment (UE) is in a good radio link condition and executes the HO at an optimal time. We analyzed the HO performance of Long-Term Evolution (LTE) and ZEUS theoretically using a geometry-based model, considering the most important HO parameter, i.e., HO margin (HOM). We derived the probabilities of HOF and PP from the analysis. The numerical results demonstrated that ZEUS HO can achieve zero HOF rate without increasing the PP rate, solving the trade-off. Furthermore, we showed that the ZEUS HO can accomplish zero HOF rate and zero PP rate simultaneously with an extension of keeping fast moving users out of small cells.

## I. INTRODUCTION

NETWORK densification is regarded as the dominant driver for wireless evolution into the era of 5G. The growth of wireless system capacity ever since the invention of the radio right up to the present can be attributed to three main factors (in decreasing order of impact): increase in the number of wireless infrastructure nodes, increased use of radio spectrum, and improvement in link efficiency [1]. Because of the projected significantly huge wireless system capacity requirement in the next ten years, the cell miniaturization and densification is touted as the most favorable way forward. Network densification through the deployment of large number of small cells is being considered as one of the most effective ways for providing increased system spectral efficiency and satisfying the explosive traffic demand [2]. It seems obvious that if we are looking for a 1000-fold increase in the system capacity, network densification through ultra-dense small cell deployments is the most appealing approach, and today's networks have already started going down this path [3].

However, in realizing the potential coverage and capacity benefits of ultra-dense small cells, operators are facing new technical challenges in mobility management, inter-cell interference coordination, and backhaul provisioning. Among these challenges, mobility management is a matter of special importance [4]. In the LTE/LTE-Advanced systems, mobility management is a key component in performance optimization where we are facing a heterogeneous network (HetNet), which is composed of different types of cells and relays. Also the high mobility requirement makes the mobility management even more challenging. Poor mobility management will result in unnecessary HOs, HOFs, and radio link failures (RLF), and hence system resources are wasted and user experiences deteriorate [5].

Due to small cell channel fading and interference, the HOF rate in HetNets is generally higher than that in macro-cell networks, and the HO from a small cell to a macro cell, i.e., pico-to-macro HO, shows the worst performance [6]. In addition, the denser the small cell deployments, the more HOFs occur owing to more pico-pico interference [7]. Moreover, if HO parameters are set such that a UE stays longer in small cells, the HOF rate is increased more [7], [8]. Furthermore, co-channel deployments of macro and small cells will result in severe interference conditions resulting in increased HOF rate compared to a homogeneous macro-cell deployment [9]. A recent LTE field test, conducted in a major city in North America, shows that the HOF problem is severe [10]. A voice over LTE call was active during a drive test to see the impact of mobility on the delay and interruption. The result shows that the HOF rate is 7.6% in urban areas and 21.7% in downtown areas.

The major cause of HOFs is a transmission failure of a handover command (HO CMD) message due to signaling in an inter-cell interference region at the cell edge, where the proportion is over 90% [11], [12]. The mobility robustness is an intricate problem because there is a trade-off between the HOF rate and the PP rate. Optimizing HO parameters to reduce HOFs would increase PPs, and vice versa [4], [6], [11].

This paragraph of the first footnote will contain the date on which you submitted your paper for review. This work was supported by 'The Cross-Ministry Giga KOREA Project' grant from the Ministry of Science, ICT and Future Planning, Korea.

Hyun-Seo Park is with Giga Communication Research Department, Electronics and Telecommunications Research Institute, Daejeon, 34129, Republic of Korea, and also with the Department of Information Communications Engineering, Chungnam National University, Daejeon, 34134, Republic of Korea (e-mail: hspark@etri.re.kr).
Yong-Seouk Choi and Tae-Joong Kim are with Giga Communication Research Department, Electronics and Telecommunications Research Institute, Daejeon, 34129, Republic of Korea (e-mail: choiys@etri.re.kr; aisma@etri.re.kr).
Byung-Chul Kim and Jae-Yong Lee are with the Department of Information Communications Engineering, Chungnam National University, Daejeon, 34134, Republic of Korea (e-mail: byckim@cnu.ac.kr; jyl@cnu.ac.kr).

*A. Literature Review*

The mobility performance has been one of the main focuses in LTE since its first release and LTE should strive to offer best mobility in a more and more challenging network [13]. Above all things, the mobility performance with small cell deployments in HetNet or ultra-dense network (UDN) is a big concern. Many literatures demonstrate that HetNet increases the probability of HOF through simulations [4], [6], [14], theoretical analysis [15], emulation with commercial smartphones [16], or field test [10], compared to homogeneous macro-cell deployment. Moreover, the higher HOF rate and the appreciably increased service interruption time in UDN are well substantiated by simulations [7], [17], [18].

Various mobility-related studies have been conducted in 3GPP. The mobility enhancements for LTE in HetNets were studied and simulation studies have shown that in the heterogeneous deployments, the mobility performance is not as good as in homogeneous deployments [6], [19]. A lot of solutions for improving the mobility robustness in LTE networks were proposed and discussed [20], [21]. Finally, some enhancements have been introduced to help improving the mobility robustness [22], [23]: UE mobility history reporting, cell-specific time-to-trigger (TTT), shorter RLF timer, i.e., T312, and context fetch. In the study for small cell enhancements [24], the mobility robustness issues in dense small cells deployments were further discussed and identified, but corresponding solutions were not specified [7].

To improve the HO performance with regard to HOF rate, UE mobility history reporting [25] increases the network's knowledge of the UE mobility speed to allow more detailed HO parameter tuning and cell-specific TTT [26] introduces target cell-specific TTT to allow network to modify the mobility event triggering per target cell, which can improve mobility robustness in HetNets. For improvements to help with recovery from RLF, the shorter RLF timer [12] allows fast recovery from RLF during the HO process and the context fetch solution [27] is that considering the case that a UE tries to access to an un-prepared target cell in HetNets, if the accessing UE's context is not available, let the target cell to get it from the source eNodeB (eNB) via X2 interface for a successful radio resource control (RRC) connection re-establishment.

HO algorithms from recent academic works, to reduce HOFs, are not much different from 3GPP works. Protecting HO CMD by using time domain or frequency domain interference coordination can decrease the HOF rate by increasing the size of HO region virtually. The mobility enhancement by control channel protection is proposed in [28]. The simulation results show that the HOF rate can be greatly reduced by muting the strongest neighboring interferer using dynamic almost-blank-subframe (ABS) coordination. The distributed HO protection resource reservation strategy for HOF reduction is proposed in [29]. For medium traffic load, the proposed reservation strategy greatly reduce the HOF rate by about 40 to 70%. However, for low or high traffic load, there is no performance gain because there is no degree of freedom to select the proper resources to avoid the interference. Moreover, it has been proven that non-ideal interference coordination among cells can lead to increased interference, which in turn results in a mobility performance degradation [6].

A detailed HO parameter tuning based on the UE mobility speed [30] and cell type [18] can decrease the HOF rate. In [30], a dynamically hybrid HOM adjusting method based on the UE speed is proposed and it reduced the HOF rate by about 40%. However, TTT is not considered in the paper and if smaller TTT to trigger the HO early is used, the HOF rate can be decreased, but the PP rate is increased. In [18], a cell specific HO parameter adaptation method is proposed and it reduced the HOF rate below 5%. However, only the given trajectory where UE moves to the center of the small cell is considered and it is a trajectory that has the smallest HOF probability.

Reference [31] introduced the dual connectivity (DC) functionality to improve user throughput and mobility performance. DC allows users to be simultaneously served by a macro and a small cell operating at different carriers. It can be generally observed that the HOF percentage is significantly lower with DC enabled. The improved HOF performance from using DC comes from always having the primary cell at the macro layer, while utilizing the small cells whenever possible for the user. However, DC does not help the HOF probability of macro cell HO and co-channel macro-pico deployment, and it has the problem of the UE complexity.

Despite a detailed HO parameter tuning based on the UE mobility speed and cell type, the HO performance in the dense intra-frequency pico-cell deployment is not acceptable even when the effect of the fast fading is not considered [32]. Therefore, to improve the HO performance, it is necessary to find a more appropriate solution beyond the adjustment of an HOM and a TTT. Considering the above reasoning, some new 3GPP Rel-13 study items on mobility enhancements for LTE have been proposed to improve the mobility robustness. Reference [13] proposed the enhancement for reliable transmission of HO signaling enabled by reception and transmission from/to multiple network nodes and potential improvements on the measurement and mobility procedures. In [33], the introduction of support for network configured UE based mobility in RRC connected mode is proposed and in [34], [35], the study of the effects of partly transferring the RRC connected mode handover and/or cell management control to the UE is proposed. The UE autonomous HO allows for a shorter HO reaction time, i.e., imply less HOFs due to long measurements and X2-like transmissions [36].

The theoretical analysis of the probabilities of HOF and PP is significant to the verification of an HO algorithm, but not much work has been done on it. That is because a theoretical analysis of the HO performance is challenging owing to the complexity of modeling the interference of the neighboring cells, and the statistics of a UE's sojourn time within a cell. In [15], the theoretical analysis of HOF and PP probabilities in HetNets, based on a simple geometric model, as a function of TTT and UE velocity was introduced. In [37], the theoretical analysis was extended, considering layer-3 (L3) filtering. In [38], the work was more extended, considering both shadowing and multi-path fading channel impairments. However, in all these works, the HOM was not considered for the simplicity.

*B. Our Previous Works*

In our previous works, we proposed the "Early Handover Preparation with Ping-Pong Avoidance" (EHOPPPA) HO that improves overall HO performance with regard to HOF rate without sacrificing PP rate and helps recovery from an RLF by itself [39], [40]. We showed that the EHOPPPA HO can achieve significant reduction in HOF rate without increasing the PP rate through simulation works [41], [42]. The simulation results showed over a 70% reduction in the HOF rate and nearly a 100% successful recovery rate from an RLF during HO without increasing the PP rate. Also, we introduced a simple geometric model for analyzing HO performance of EHOPPPA HO where the HOM was not considered for the simplicity, comparing with the LTE HO [15]. Through preliminary theoretical analysis, we demonstrated that the EHOPPPA HO can solve the trade-off, achieving zero HOF rate without increasing the PP rate, regardless of the UE velocity.

*C. Contributions*

We renamed EHOPPPA as ZEUS, which is much easier to pronounce and well represents the strengths that ZEUS HO can achieve zero HOF regardless of the UE velocity and the size of HO region. The main goal of this paper is to introduce a geometric model considering the HOM, the most important HO parameter, for analyzing HO performance of LTE HO and ZEUS HO and demonstrate that ZEUS HO can solve the trade-off. Moreover, we show the ZEUS HO can achieve zero HOF rate with the lowest PP rate, regardless of the size of HO region as well as the UE velocity. In [43], it is said that continuing on HO parameter tuning based on the UE mobility for enhancing event triggering seems unlikely to be able to handle a large diversity of deployments well and turn out to be counter-productive. The mobility enhancement not based on UE speed but on radio conditions themselves, is more intuitive, and the ZEUS HO has the UE take action considering real observed radio conditions.

In this paper, extending our prior works, we
- explain the ZEUS HO in more detail;
- provide the geometry-based HO models for LTE and ZEUS, considering the HOM. To the best of our knowledge, there are no analytical results that study HO performance in LTE HetNets, considering the HOM;
- derive the probabilities of HOF and PP in closed-form expressions from the HO performance analysis;
- calculate the numerical results of those in three macro-pico distance cases as a function of UE velocity, comparing the HO performance of LTE and ZEUS;
- demonstrate that the ZEUS HO can solve the trade-off, achieving zero HOF rate without increasing the PP rate, based on the numerical results, regardless of the HO models;
- show that the ZEUS HO can achieve zero HOF and zero PP rate simultaneously with an extension of keeping fast moving users out of small cells.

*D. Organization of the Paper*

The rest of the paper is organized as follows. Section II discusses the LTE HO and its problem with mobility robustness, and presents the ZEUS HO to resolve the problem. In Section III, the analysis models for HO performance, comparing LTE with ZEUS, are introduced and the probabilities of HOF and PP are derived from the analysis. In Section IV, the numerical results of the probabilities of HOF and PP as a function of UE velocity are showed and the significant findings drawn from the results are discussed. Finally, Section V offers the concluding remarks.

## II. LTE HANDOVER AND ZEUS HANDOVER

In this section, we discuss the HO procedure in LTE networks and its performance problem with mobility robustness. Then, we present the ZEUS HO to resolve the problem.

*A. LTE Handover*

In LTE networks, mobile-assisted, network-controlled HOs are performed, as shown in Fig. 1 [42], [44], [45]. The UE conducts HO measurements and processing. HO measurements are usually based on downlink (DL) reference signal received power (RSRP) estimations, while the processing takes place to filter out the effects of fading and estimation imperfections in the HO measurements. After processing, if a certain HO event occurs according to the filtered measurements, the UE sends a measurement report (MR) message to the source eNB (S-eNB). When the radio signal of a neighbor cell is better than that of the serving cell by a specified HOM offset, that is, an A3 event defined in [46] is met, a TTT timer is initiated. If such a state remains in existence throughout the duration of the TTT period, then, usually, an HO event is triggered upon conclusion of this period.

The HO preparation phase then starts when the S-eNB issues a handover request message to the target eNB (T-eNB), which carries out admission control according to the quality of service requirement of the UE. After the admission, the T-eNB prepares the HO process, and sends a handover request ACK message to the S-eNB. After receiving the handover request ACK message, S-eNB usually stops DL data transmission to the UE and starts data forwarding to the T-eNB, and transmits

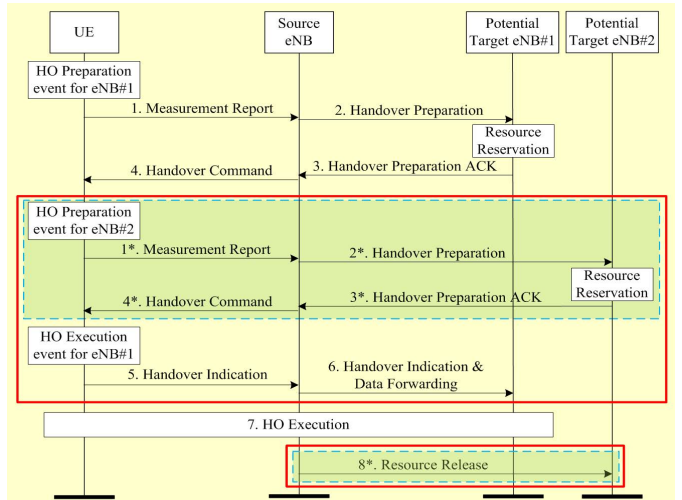

Fig. 1. The procedure of LTE and ZEUS handover (procedure in red box is only applicable to ZEUS handover and procedure in blue box is optional and exists only in case of multiple handover preparations).

an HO CMD to the UE.

Immediately after receiving an HO CMD, during the HO execution and completion phase, the UE synchronizes with the target cell and accesses it. The UE sends a handover complete message to the T-eNB when the HO procedure is finished. The T-eNB, which can then start transmitting data to the UE, sends a path switch request message to inform the network that the UE has changed its serving cell. Thereafter, the network switches the DL data path from the S-eNB to the T-eNB.

During the HO process, an RLF occurs frequently due to a DL physical layer failure caused by DL interference from the neighbor cells. The UE may declare an RLF in a number of scenarios including the following: a timer T310 expiry after a DL physical layer failure when the block error rate of the physical downlink control channel (PDCCH) is greater than 10%, random access problems, maximum radio link control (RLC) retransmissions, or an HO failure. Once an RLF is declared, the UE begins an RLF recovery procedure. The UE attempts a cell selection and a connection re-establishment procedure with the selected cell. The re-establishment procedure succeeds only if the UE selects a cell of the S-eNB or a prepared T-eNB. If the procedure fails, then the UE enters into idle mode and attempts non-access stratum (NAS) recovery [42], [44]–[46]. The duration of service interruption is reported to be about 80 ms to 130 ms in a successful HO, 800 ms to 3,000 ms in RLF recovery after an HOF, and 3,000 ms to 5,000 ms in NAS recovery after an RLF recovery failure [10].

*B. Problem Formulation of LTE Handover Performance*

The HO parameter tuning has been mainly studied and used to improve the HO performance in LTE. Table I shows the HO parameter sets which are used for LTE HetNet HO performance simulation calibration in 3GPP excerpted from Table 5.3.2.1 of [6]. The HO parameters should be properly set to avoid HOFs. However, there is a trade-off between the HOF rate and the PP rate. Fig. 2 shows average HOF rate and PP rate curves from the 3GPP LTE HetNet HO performance simulation calibration results, excerpted from Fig. 5.5.1.2.3 and Fig. 5.5.1.3.1 of [6]. The simulation assumptions, the detailed parameters, and the considered scenario can be found in clause 5.2 and 5.3 of [6]. The HO parameters in 'Set 1' delay an HO until the target cell is better enough than the serving cell and trigger the HO late. In this case, the PP rate can be decreased, but the HOF rate is increased as shown in Fig. 2. The 'Too Late HO' failure [44] can be occurred where an RLF occurs in the source cell before the HO is triggered or during the HO procedure and the UE attempts to re-establish the radio link connection with the target cell. On the other hand, if we select the HO parameters in 'Set 5' to trigger the HO early, the HOF rate can be decreased, but the PP rate is increased. In this case, the 'Too Early HO' failure [44] can be occurred where an RLF occurs shortly after a successful HO from a source cell to a target cell or during the HO procedure and the UE attempts to re-establish the radio link connection with the source cell.

TABLE I
CONFIGURATION PARAMETER SETS FOR HANDOVER PERFORMANCE SIMULATION CALIBRATION [6]

| Profile | Set 1 | Set 2 | Set 3 | Set 4 | Set 5 |
|---|---|---|---|---|---|
| UE speed [Km/h] | {3, 30, 60, 120} | | | | |
| Cell Loading [%] | 100 | | | | |
| TTT [ms] | 480 | 160 | 160 | 80 | 40 |
| A3 offset [dB] | 3 | 3 | 2 | 1 | -1 |
| L1 to L3 period [ms] | 200 | | | | |
| RSRP L3 Filter K | 4 | 4 | 1 | 1 | 0 |

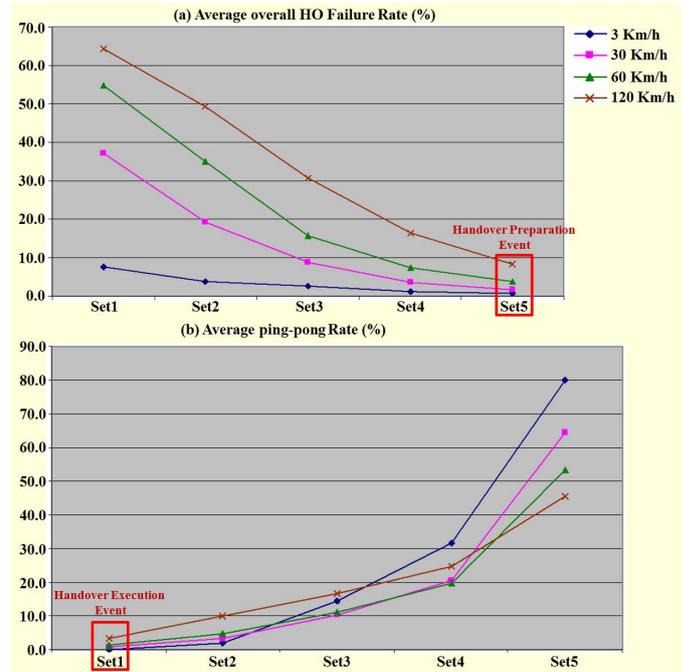

Fig. 2. (a) Average overall handover failure rate curves and (b) average ping-pong rate curves [6], and an example of handover preparation/execution events in ZEUS handover (two red rectangles).

The mobility management has been and continues to be a cornerstone in LTE. However, the introduction of small cells has been bringing a more serious problem in mobility robustness. The major cause of HOFs is a transmission failure of HO CMD due to signaling in an inter-cell interference region at the cell edge. The robust handover signaling has become an intricate problem in various cell border situations as seen from real network deployments and operators have made strenuous efforts in the network tuning and HO parameter adjustments. The new wireless communication trends, such as ultra-dense networks, an extreme beamforming, a higher frequency, as well as non-ideal real network deployments, are expected to make the mobility robustness problem far more serious.

*C. ZEUS Handover*

The rationale behind the ZEUS HO [39]–[42] is very intuitive. The ZEUS HO assures that the HO signaling is completed robustly while a UE is in a good radio link condition with the serving cell, and the HO is executed when the radio signal of a target cell is better enough than that of the serving cell. The ZEUS HO splits an HO event into an HO preparation (HOP) event and an HO execution (HOE) event. The HOP event is used for an 'Early Handover Preparation' and the HOE event is used for an HO execution with 'Ping-Pong Avoidance'. Therefore, the former name of ZEUS was EHOPPPA. If an HOP event such as 'Set 5' is chosen, then an early HO preparation triggered by this event can decrease the HOF rate. And, if an HOE event such as 'Set 1' is chosen, then the HO execution triggered by this event can prevent PP occurrences from being accompanied by the above premature HO as shown in Fig. 2.

In the ZEUS HO, mobile-assisted, network-controlled HOs are applied in the same manner as in the LTE HO. However, while the LTE HO is fully network-controlled, the ZEUS HO is hybrid-controlled in that it transfers part of the control of the cell selection at an HO to a UE. The ZEUS HO consists of network-controlled HO preparation and mobile-controlled HO execution. With the ZEUS HO, a UE backs up one or more 'early' HO CMDs and executes an HO to an optimal target cell selected among multiple prepared candidate target cells based on the backed-up 'early' HO CMDs, at an optimal time.

The ZEUS HO procedure is similar to that of the LTE HO except procedure in the red box, as shown in Fig. 1. When an HOP event is triggered, the UE sends an MR to the S-eNB. The HOP event can be an A3 event with offset1, for example. The S-eNB perfoms an HO preparation to a potential T-eNB based on the MR. The potential T-eNB performs admission control and resource reservation, and sends a handover request ACK to the S-eNB. The S-eNB sends an 'early' HO CMD to the UE.

The ZEUS HO supports the feature of multiple HO preparations inherently and gives a cell selection opportunity to the UE based on multiple HO preparations. If an HOP event for another potential T-eNB is triggered, then another HO preparation can be performed (procedure from 1* to 4* in blue box in Fig. 1). The LTE HO also supports the feature of multiple HO preparations where the S-eNB is allowed to perform HO preparations with multiple T-eNBs [45]. However, multiple HO preparations are helpful for only a successful re-establishment after an RLF occurs during the HO procedure. Although the S-eNB is allowed to perform HO preparations with multiple T-eNBs, a UE can receive only one HO CMD for a prepared T-eNB selected by the S-eNB. Thus, the HO is successful only if the UE accesses that target cell. Therefore, the gain from this feature is too limited because it does not do much for the HO itself. But, in the ZEUS HO, multiple HO preparations are helpful for a successful HO as well as a successful re-establishment.

In the ZEUS HO, after receiving an 'early' HO CMD, the UE does not execute an HO immediately, unlike in the LTE HO, but simply backs up 'early' HO CMDs and performs measurements continually. Then, the UE determines an optimal time of HO execution and an optimal target cell based on the continual measurements. Because the UE obtains the best knowledge regarding its own radio conditions in a timely manner, its decision can be the optimum. Once the UE determines an optimal time of HO execution and an optimal target cell which is triggered by an HOE event, the UE sends a handover indication, notifying the S-eNB of an immediate HO execution and the selected T-eNB. An HOE event can be an A3 event with offset2, where offset2 is bigger than offset1. Then the procedures in the HO execution and completion phase are performed, as in the LTE HO. Even if the transmission of handover indication fails, the UE can execute an HO successfully to the selected T-eNB because it has already received 'early' HO CMD for that eNB.

Usually, in the LTE HO, after an HO event entering condition is met, a UE waits additionally for a TTT in order to avoid a premature HO initiation. However, a TTT causes an extra HO delay and is one reason for an increase of the HOF rate in HetNets [32]. Moreover, mobility speed estimation (MSE)-based TTT scaling does not work well because the MSE itself is not accurate in a HetNet environment, and it is hard to adjust a TTT in the real network configurations. Normally, in the ZEUS HO, an HOP event and an HOE event do not use a TTT. Instead, a suppositional 'time-to-execute' (TTE), which is the elapsed time from a receipt of an 'early' HO CMD to an HO execution, is unforced and automatically well scaled depending on the mobility speed of the UE and real network configurations. This is explained in more detail in Section IV.

III. HANDOVER MODELS AND ANALYSIS

In this section, we discuss a conventional LTE HetNet HO model and present an LTE HO model and a ZEUS HO model, considering the HOM. Then, we derive the HO performance metrics such as HOF and PP probabilities of LTE and ZEUS HO through the theoretical analysis.

A. Conventional LTE HetNet Handover Model

A theoretical analysis of the HO performance is challenging due to the complexity of modeling the interference of the neighboring cells, and the statistics of a UE's sojourn time within a cell. Many literatures adopt a geometry-based model in the theoretical analysis of HOFs and PPs [15], [37], [38], [42]. Fig. 3 [15] illustrates an example for HO trigger locations and HOF locations generated by a simulator that implements 3GPP LTE HetNet simulation assumptions. The HO trigger locations are observed to be scattered around the perimeter of a circle. On the contrary, the HOF locations resemble a deformed circle, where deforming occurs due to sector antennas implemented at the eNBs. If we neglect the impact of sectorized cell structure, the HOF locations can be approximated by a circle [38]. Thus, it is reasonable to model the HO trigger locations and HOF locations geometrically as concentric circles [15], [37], [38]. The conventional LTE HetNet HO model in [15], [37], [38] has three concentric circles which are an HOF circle for a macro cell UE (MUE), an HOF circle for a pico cell UE (PUE), and an HO trigger circle which is the same as the pico cell coverage circle because an HOM of 0 dB is assumed for the simplicity. The notations used in this section are shown in Table II.

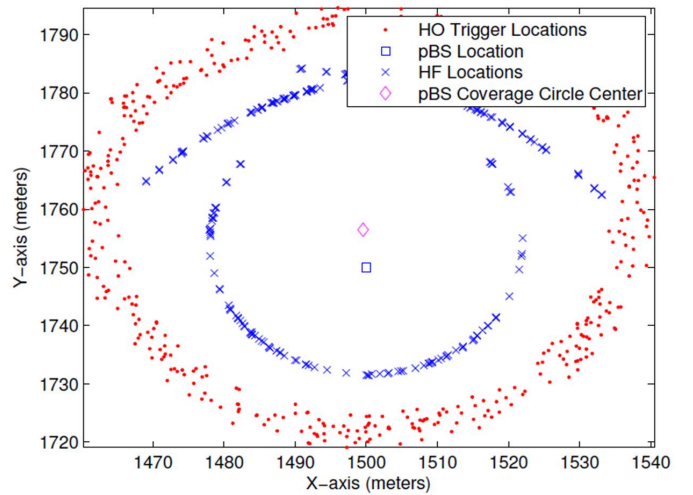

Fig. 3. Handover trigger locations and handover failure locations generated by a simulator implementing LTE HetNet simulation assumptions [15].

TABLE II
NOTATIONS

| Notation | Explanation |
|---|---|
| $v$ | The velocity of the UE |
| $T_m$ | The value of TTT for macro to pico handover |
| $T_p$ | The value of TTT for pico to macro handover |
| $T_{pp}$ | A pre-determined minimum time-of-stay |
| $R$ | The radius of the pico cell coverage circle |
| $r_{mp}$ | The radius of the HOM circle for MUE |
| $r_{me}$ | The radius of the HOE circle for MUE (only in ZEUS) |
| $r_m$ | The radius of the HOF circle for MUE |
| $r_{pp}$ | The radius of the HOM circle for PUE |
| $r_{pe}$ | The radius of the HOE circle for PUE (only in ZEUS) |
| $r_p$ | The radius of the HOF circle for PUE |

### B. LTE Handover Model and Analysis

Fig. 4 illustrates an LTE HO model, considering the HOM. A UE starts as an MUE and moves along a straight line toward an arbitrary direction. The MUE becomes a PUE if it is successfully handed over to the pico cell, and becomes an MUE again if it is successfully handed over to the macro cell.

An MUE starts at the pico cell coverage circle, e.g., the point A, and moves along a straight line toward an arbitrary direction. As soon as an MUE enters the MUE HOM circle, a TTT of duration $T_m$ is initiated. After the TTT is triggered, the MUE does not make an HO into the pico eNB (PNB) if it leaves the MUE HOM circle before the end of the TTT. If the TTT is expired before it leaves the MUE HOM circle, the MUE makes an HO into the PNB. An MUE HOF occurs if the distance $vT_m$ travelled by the MUE during the TTT is larger than the distance between the location where the MUE trajectory intersects with the MUE HOM circle and the location where the MUE trajectory intersects with the MUE HOF circle.

As soon as a PUE enters the PUE HOM circle, a TTT of duration $T_p$ is initiated. If the TTT is expired, the PUE makes an HO into the macro eNB (MNB). A PUE HOF occurs if the distance $vT_p$ travelled by the PUE during the TTT is larger than the distance between the location where the PUE trajectory intersects with the PUE HOM circle and the location where the PUE trajectory intersects with the PUE HOF circle. An HO from macro cell to pico cell, then HO back to macro cell is defined as a PP if the time-of-stay (ToS) connected in pico cell is less than a pre-determined minimum ToS, i.e., $T_{pp}$ time unit, where $T_{pp}$ is 1 s as defined by 3GPP in clause 5.2.2 of [6].

The radius of the pico cell coverage circle is denoted by $R$, the radii of the HOM circles for MUE and PUE are denoted by $r_{mp}$ and $r_{pp}$, and the radii of the HOF circles for MUE and PUE are denoted by $r_m$ and $r_p$, respectively, where $r_m < r_{mp} < R < r_{pp} < r_p$.

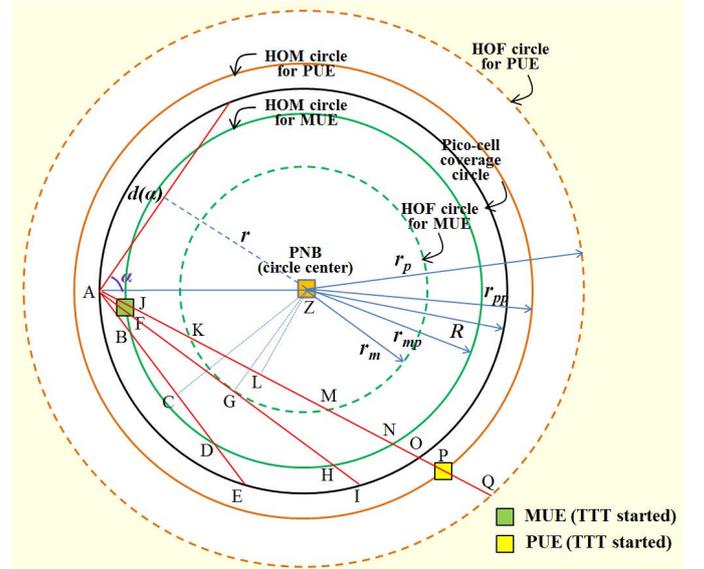

Fig. 4. Analysis model of macro UE and pico UE handover with standard LTE handover.

#### 1) The probability of a random chord

Bertrand's paradox [47] is used for the theoretical analysis of the HO performance. Bertrand's paradox aims to find the probability of a random chord of a circle with a radius $R$ being larger than a threshold. Let $d(\alpha) = 2R\cos(\alpha)$ denote the length of the chord determined by the intersection points between an MUE trajectory and the pico cell coverage circle; $v$ be the velocity of the UE on this chord; $\alpha$ denote the angle of the chord, that is, the UE trajectory, with respect to the horizontal axis; and $r$ be the minimum distance from the center of the pico cell coverage circle to the trajectory of the MUE. The probability density function of $r = \sqrt{R^2 - d(\alpha)^2/4}$ for $\alpha$ is given by

$$f(r) = \frac{2}{\pi\sqrt{R^2 - r^2}} \quad (1)$$

Based on (1), for two arbitrary chord lengths $d_1$ and $d_2$, with $d_1 \leq d_2$, the probability of $d(\alpha)$ being between $d_1$ and $d_2$ is

$$P(d_1 < d(\alpha) < d_2) = \frac{2}{\pi} \tan^{-1} \frac{r}{\sqrt{R^2 - r^2}} \Big|_{\sqrt{R^2 - d_2^2/4}}^{\sqrt{R^2 - d_1^2/4}} \quad (2)$$

and is used to calculate the probabilities of the HOF and PP as stated in [15].

#### 2) Probability of No Handover for MUEs

As soon as an MUE enters the MUE HOM circle, a TTT of duration $T_m$ is initiated. After the TTT is triggered, the MUE does not make an HO into the PNB if it leaves the MUE HOM circle before the end of the TTT. This is the case where the distance $vT_m$ travelled by the MUE during the TTT is larger than the distance between the location where the MUE trajectory intersects the MUE HOM circle and the location where the MUE trajectory intersects the MUE HOM circle once again. And the chord should not intersect with the MUE HOF circle because it is the case where there is an MUE HOF. As a consequence, the no HO (NHO) probability can be expressed as follows

$$P_{\text{NHO}} = \begin{cases} P\left(d(\alpha) < 2\sqrt{R^2 - r_m^2}\right), & \text{if } vT_m > 2\sqrt{r_{mp}^2 - r_m^2} \\ P\left(d(\alpha) < 2R\cos\theta\right), & \text{if } vT_m < 2\sqrt{r_{mp}^2 - r_m^2} \end{cases} \quad (3)$$

where $2\sqrt{R^2 - r_m^2}$ is the chord length when the UE's trajectory is tangent to the MUE HOF circle, i.e., the segment AI perpendicular to the segment GZ.

In (3), $\theta$ is the value of $\alpha$, where the distance between the location where the MUE trajectory intersects the MUE HOM circle and the location where the MUE trajectory intersects the MUE HOM circle once again is equal to $vT_m$. The chord is the segment AE perpendicular to the segment CZ, where the length of the segment BD is $vT_m$.

$$\{d(\alpha) = 2R\cos\theta\} = \{|BD| = vT_m\} \quad (4)$$

In (4), $\theta$ can be expressed as

$$2\sqrt{r_{mp}^2 - R^2 \sin^2\theta} = vT_m \quad (5)$$

After some manipulations, $\theta$ can be derived as

$$\theta = \sin^{-1}\sqrt{\left(\frac{r_{mp}}{R}\right)^2 - \left(\frac{vT_m}{2R}\right)^2} \quad (6)$$

Using (2), (3), and (6), the NHO probability can be written as

$$P_{\text{NHO}} = \begin{cases} 1 - \frac{2}{\pi}\tan^{-1}\frac{r_m}{\sqrt{R^2 - r_m^2}}, & \text{if } vT_m > 2\sqrt{r_{mp}^2 - r_m^2} \\ 1 - \frac{2}{\pi}\theta, & \text{if } vT_m < 2\sqrt{r_{mp}^2 - r_m^2} \end{cases} \quad (7)$$

*3) Probability of Handover Failure of MUEs*

If $d(\alpha) \geq 2\sqrt{R^2 - r_m^2}$, the MUE trajectory intersects with the MUE HOF circle, and thus the possibility of MUE HOF exists. An MUE HOF occurs if the distance $vT_m$ travelled by the MUE during the TTT is larger than the distance between the location where the MUE trajectory intersects the MUE HOM circle and the location where the MUE trajectory intersects the MUE HOF circle. Note that for the special case, MUE HOF is not observed if $vT_m < r_{mp} - r_m$. As a consequence, the MUE HOF probability can be expressed as follows

$$P_{\text{HF,m}} = \begin{cases} P\left(d(\alpha) > 2\sqrt{R^2 - r_m^2}\right), & \text{if } vT_m > \sqrt{r_{mp}^2 - r_m^2} \\ 0, & \text{if } vT_m < r_{mp} - r_m \\ P\left(d(\alpha) > 2R\cos\beta\right), & \text{otherwise} \end{cases} \quad (8)$$

In (8), $\beta$ is the value of $\alpha$, where the distance between the location where the MUE trajectory intersects the MUE HOM circle and the location where the MUE trajectory intersects the MUE HOF circle is equal to $vT_m$. The chord is the segment AO perpendicular to the segment LZ, where the length of the segment JK is equal to $vT_m$.

$$\{d(\alpha) = 2R\cos\beta\} = \{|JK| = vT_m\} \quad (9)$$

The length of the segment JK is

$$|JK| = |JL| - |KL| = vT_m \quad (10)$$

In (9), using (10), $\beta$ can be expressed as

$$\sqrt{r_{mp}^2 - R^2 \sin^2\beta} - \sqrt{r_m^2 - R^2 \sin^2\beta} = vT_m \quad (11)$$

After some manipulations, $\beta$ can be derived as

$$\beta = \sin^{-1}\sqrt{\left(\frac{r_m}{R}\right)^2 - \left(\frac{r_{mp}^2 - r_m^2 - (vT_m)^2}{2RvT_m}\right)^2} \quad (12)$$

Using (2), (8), and (12), the MUE HOF probability can be written as

$$P_{\text{HF,m}} = \begin{cases} \frac{2}{\pi}\tan^{-1}\frac{r_m}{\sqrt{R^2 - r_m^2}}, & \text{if } vT_m > \sqrt{r_{mp}^2 - r_m^2} \\ 0, & \text{if } vT_m < r_{mp} - r_m \\ \frac{2}{\pi}\beta, & \text{otherwise} \end{cases} \quad (13)$$

Complying with the definition of HOF rate defined by 3GPP in [6] as stated in clause 5.2.1.3, the MUE HOF probability can be redefined as

$$P_{\text{HF,m,3GPP}} = \frac{P_{\text{HF,m}}}{1 - P_{\text{NHO}}} \quad (14)$$

*4) Probability of Handover Failure of PUEs*

A PUE HOF occurs if the distance $vT_p$ travelled by the PUE during the TTT is larger than the distance between the location where the PUE trajectory intersects the PUE HOM circle and the location where the PUE trajectory intersects the PUE HOF circle. Note that for the special case, PUE HOF is not observed if $vT_p < r_p - r_{pp}$. Also, note that for the special case, either PUE HO or PUE HOF is not observed if $vT_m > 2\sqrt{r_{mp}^2 - r_m^2}$, either due to NHO, or due to MUE HOF. So, the PUE HOF probability can be expressed as follows,

$$P_{\text{HF,p}} = \begin{cases} 0, & \text{if } vT_m > 2\sqrt{r_{mp}^2 - r_m^2} \\ 0, & \text{if } vT_p < r_p - r_{pp} \\ P\left(\max(d_{HF,p}, d_{NHO}) < d(\alpha) < d_{HF,m}\right), & \text{otherwise} \end{cases} \quad (15)$$

where $d_{HF,p}$ is the minimum of $d(\alpha)$ when the PUE HOF occurs, $d_{NHO}$ is the maximum of $d(\alpha)$ when the NHO occurs, and $d_{HF,m}$ is the minimum of $d(\alpha)$ when the MUE HOF occurs.

In (15), let $d_{HF,p} = 2R\cos\delta$, and $\delta$ is the value of $\alpha$, where the distance between the location where the PUE trajectory intersects the PUE HOM circle and the location where the PUE trajectory intersects the PUE HOF circle is equal to $vT_p$. The chord is the segment AO perpendicular to the segment LZ, where the length of the segment PQ is equal to $vT_p$.

$$\{d(\alpha) = 2R\cos\delta\} = \{|PQ| = vT_p\} \quad (16)$$

In (16), $\delta$ can be expressed as

$$r_p^2 = R^2 \sin^2\delta + \left(\sqrt{r_{pp}^2 - R^2 \sin^2\delta} + vT_p\right)^2 \quad (17)$$

After some manipulations, $\delta$ can be derived as

$$\delta = \sin^{-1}\sqrt{\left(\frac{r_{pp}}{R}\right)^2 - \left(\frac{r_p^2 - r_{pp}^2 - (vT_p)^2}{2RvT_p}\right)^2} \quad (18)$$

Using (2), (3), (8), and (15), the PUE HOF probability can be written as

$$P_{\text{HF,p}} = \begin{cases} 0, & \text{if } vT_m > 2\sqrt{r_{mp}^2 - r_m^2} \\ 0, & \text{if } vT_p < r_p - r_{pp} \\ P(A < d(\alpha)), & \text{if } vT_m < r_{mp} - r_m \\ P(A < d(\alpha) < 2\sqrt{R^2 - r_m^2}), & \text{if } vT_m > \sqrt{r_{mp}^2 - r_m^2} \\ P(A < d(\alpha) < 2R\cos\beta), & \text{otherwise} \end{cases} \quad (19)$$

, where $A = \max(2R\cos\delta, 2R\cos\theta)$

Complying with the definition of HOF rate defined by 3GPP, the PUE HOF probability can be redefined as

$$P_{\text{HF,p,3GPP}} = \frac{P_{\text{HF,p}}}{1 - P_{\text{NHO}} - P_{\text{HF,m}}} \quad (20)$$

*5) Probability of Ping-Pong*

A PP occurs when a PUE stays less than $T_{pp}$ time unit within the pico cell, where $T_{pp}$ is 1 s. Note that for the special case where either PUE HO or PUE HOF is not observed if $vT_m > 2\sqrt{r_{mp}^2 - r_m^2}$, either due to NHO, or due to MUE HOF. Also, note that for the special case where PP is not observed if the minimum traveling distance during ToS in pico cell is greater than $vT_{pp}$. As a consequence, the PP probability can be expressed as follows,

$$P_{\text{PP}} = \begin{cases} 0, & \text{if } vT_m > 2\sqrt{r_{mp}^2 - r_m^2} \\ 0, & \text{if } \ell_{\min}(ToS_{pico}) > vT_{pp} \\ P(d_{NHO} < d(\alpha) < \min(d_{PP}, d_{HF,m})), & \text{otherwise} \end{cases} \quad (21)$$

where $d_{PP}$ is the maximum of $d(\alpha)$ when the PP occurs, $d_{NHO}$ is the maximum of $d(\alpha)$ when the NHO occurs, and $d_{HF,m}$ is the minimum of $d(\alpha)$ when the MUE HOF occurs. We regard the case, where a PUE HOF occurs and ToS in pico cell is less than $T_{pp}$, as a PP also.

In (21), the minimum traveling distance during ToS in pico cell in the second special case is {(the length of the segment NP) + $vT_p$}. The chord is the segment AO perpendicular to the segment LZ, where the length of the segment JN is $vT_m$.

$$\ell_{\min}(ToS_{pico}) = |NP| + vT_p, \text{where } |JN| = vT_m \quad (22)$$

In (22), $l_{\min}(ToS_{pico})$ can be expressed as

$$\ell_{\min}(ToS_{pico}) = \sqrt{r_{pp}^2 - \left(\sqrt{r_{mp}^2 - (\frac{vT_m}{2})^2}\right)^2} - \frac{vT_m}{2} + vT_P \quad (23)$$

In (21), let $d_{PP} = 2R\cos\gamma$, where $\gamma$ is the value of $\alpha$, where the distance between the location where the MUE makes an HO into the PNB and the location where the PUE makes an HO into the MNB is equal to $vT_{pp}$. The chord is the segment AO perpendicular to the segment LZ, where the length of {(the segment AO) − (the segment AJ + $vT_m$) + (the segment OP + $vT_p$)} is equal to $vT_{pp}$.

$$\begin{cases} d(\alpha) = 2R\cos\gamma \\ d(\alpha) - (|AJ| + vT_m) + (|OP| + vT_p) = vT_{pp} \end{cases} \quad (24)$$

In (24), the length of the segment AJ is

$$|AJ| = |AL| - |JL| = R\cos\gamma - \sqrt{r_{mp}^2 - R^2\sin^2\gamma} \quad (25)$$

In (24), the length of the segment OP is

$$|OP| = |LP| - |LO| = \sqrt{r_{pp}^2 - R^2\sin^2\gamma} - R\cos\gamma \quad (26)$$

Using (24), (25), and (26), $\gamma$ can be expressed as

$$\sqrt{r_{mp}^2 - R^2\sin^2\gamma} - vT_m + \sqrt{r_{pp}^2 - R^2\sin^2\gamma} + vT_p = vT_{pp} \quad (27)$$

After some manipulations, $\gamma$ can be derived as

$$\gamma = \sin^{-1}\sqrt{\left(\frac{r_{pp}}{R}\right)^2 - \left(\frac{X^2 + r_{pp}^2 - r_{mp}^2}{2RX}\right)^2}, \quad (28)$$

where $X = (vT_m - vT_p + vT_{pp})$

Using (2), (3), (8), and (21), the PP probability can be written as

$$P_{\text{PP}} = \begin{cases} 0, & \text{if } vT_m > 2\sqrt{r_{mp}^2 - r_m^2} \\ 0, & \text{if } \ell_{\min}(ToS_{pico}) > vT_{pp} \\ P(2R\cos\theta < d(\alpha) < 2R\cos\gamma), & \text{if } vT_m < r_{mp} - r_m \\ P(2R\cos\theta < d(\alpha) < B), & \text{if } vT_m > \sqrt{r_{mp}^2 - r_m^2} \\ P(2R\cos\theta < d(\alpha) < C), & \text{otherwise} \end{cases} \quad (29)$$

, where $B = \min(2R\cos\gamma, 2\sqrt{R^2 - r_m^2})$,
$C = \min(2R\cos\gamma, 2R\cos\beta)$

Complying with the definition of PP rate defined by 3GPP in [6] as stated in clause 5.2.2, the PP probability can be redefined as

$$P_{\text{PP,3GPP}} = \frac{P_{\text{PP}}}{1 - P_{\text{NHO}} - P_{\text{HF,m}}} \quad (30)$$

*C. ZEUS Handover Model and Analysis*

Fig. 5 illustrates a ZEUS HO model, considering the HOM. An MUE starts at the pico cell coverage circle, e.g., the point A, and moves along a straight line toward an arbitrary direction. As soon as an MUE enters the MUE HOP circle, an HO preparation is initiated. The MUE does not make an HO into the PNB if it leaves the MUE HOP circle before the MUE trajectory intersects with the MUE HOE circle. If the MUE trajectory intersects with the MUE HOE circle, then the MUE makes an HO into the PNB. An MUE HOF does not occur because the MUE trajectory always intersects with the MUE HOE circle before it intersects with the MUE HOF circle.

Fig. 5. Analysis model of macro UE and pico UE handover with ZEUS handover.

As soon as a PUE enters the PUE HOP circle, an HO preparation is initiated. If the PUE trajectory intersects with the PUE HOE circle, then the PUE makes an HO into the MNB. Likewise, a PUE HOF does not occur because the PUE trajectory always intersects with the PUE HOE circle before it intersects with the PUE HOF circle.

The radius of the pico cell coverage circle is denoted by $R$, the radii of the HOP circles for the MUEs and PUEs are denoted by $r_{mp}$ and $r_{pp}$, and the radii of the HOF circles for the MUEs and PUEs are denoted by $r_m$ and $r_p$, respectively, as in LTE case. And, the radii of the HOE circles for the MUEs and PUEs are denoted by $r_{me}$ and $r_{pe}$, respectively, where $r_m < r_{me} < r_{mp} < R < r_{pp} < r_{pe} < r_p$.

*1) Probability of No Handover for MUEs*

As soon as an MUE enters the MUE HOP circle, an HO preparation is initiated. The MUE does not make an HO into the PNB if it leaves the MUE HOP circle before the MUE trajectory intersects with the MUE HOE circle. As a consequence, the NHO probability can be expressed as follows

$$P_{\text{NHO}} = P\left(d(\alpha) < 2\sqrt{R^2 - r_{me}^2}\right) \quad (31)$$

where $2\sqrt{R^2 - r_{me}^2}$ is the chord length when the UE's trajectory is tangent to the MUE HOE circle, i.e., the segment AC perpendicular to the segment BZ.

Using (2), and (31), the NHO probability can be written as

$$P_{\text{NHO}} = 1 - \frac{2}{\pi} \tan^{-1} \frac{r_{me}}{\sqrt{R^2 - r_{me}^2}} \quad (32)$$

*2) Probability of Extra Handover Preparation for MUEs*

The cost of the ZEUS HO is an extra HO preparation (EHOP), which means that the prepared cell is not used for either an HO or an RLF recovery. The HO preparation is not initiated if an MUE leaves the pico cell coverage circle before the MUE trajectory intersects with the MUE HOP circle. As a consequence, the no HO preparation (NHOP) probability can be expressed as follows

$$P_{\text{NHOP}} = P\left(d(\alpha) < 2\sqrt{R^2 - r_{mp}^2}\right) \quad (33)$$

where $2\sqrt{R^2 - r_{mp}^2}$ is the chord length when the UE's trajectory is tangent to the MUE HOP circle, i.e., the segment AG perpendicular to the segment FZ.

Using (2), and (33), the NHOP probability can be written as

$$P_{\text{NHOP}} = 1 - \frac{2}{\pi} \tan^{-1} \frac{r_{mp}}{\sqrt{R^2 - r_{mp}^2}} \quad (34)$$

Therefore, the EHOP probability can be expressed as follows

$$P_{\text{EHOP}} = P_{\text{NHO}} - P_{\text{NHOP}} \quad (35)$$

*3) Probability of Handover Failure of MUEs*

An MUE HOF does not occur, because the MUE trajectory always intersects with the MUE HOE circle before it intersects with the MUE HOF circle. Therefore, the MUE HOF probability is

$$P_{\text{HF,m}} = 0, \quad \text{if } r_m < r_{me} < r_{mp} \quad (36)$$

*4) Probability of Handover Failure of PUEs*

A PUE HOF does not occur, because the PUE trajectory always intersects with the PUE HOE circle before it intersects with the PUE HOF circle. Therefore, the PUE HOF probability is

$$P_{\text{HF,p}} = 0, \quad \text{if } r_{pp} < r_{pe} < r_p \quad (37)$$

*5) Probability of Ping-Pong*

A PP occurs when a PUE stays less than $T_{pp}$ time unit within the pico cell, where $T_{pp}$ is 1 s. Note that for the special case where PP is not observed if the minimum traveling distance during ToS in pico cell is greater than $vT_{pp}$. Therefore, the PP probability can be expressed as follows,

$$P_{\text{PP}} = \begin{cases} 0, & \text{if } \sqrt{r_{pe}^2 - r_{me}^2} > vT_{pp} \\ P\left(d_{NHO} < d(\alpha) < 2R\right), & \text{if } (r_{me} + r_{pe}) < vT_{pp} \\ P\left(d_{NHO} < d(\alpha) < 2R\cos\phi\right), & \text{otherwise} \end{cases} \quad (38)$$

where $d_{NHO}$ is the maximum of $d(\alpha)$ when the NHO occurs, and $\sqrt{r_{pe}^2 - r_{me}^2}$ is the minimum traveling distance during ToS in pico cell, i.e., the length of the segment BE, when the UE's trajectory is tangent to the MUE HOE circle, i.e., the segment AC perpendicular to the segment BZ.

In (38), $\phi$ is the value of $\alpha$, where the distance between the location where the MUE makes an HO into the PNB and the location where the PUE makes an HO into the MNB is equal to $vT_{pp}$. The chord is the segment AO perpendicular to the segment KZ, where the length of the segment IQ is $vT_{pp}$.

$$\{d(\alpha) = 2R\cos\phi\} = \{|IQ| = vT_{pp}\} \quad (39)$$

In (39), the length of the segment IQ is

$$|IQ| = |IK| + |KQ| = \sqrt{r_{me}^2 - R^2 \sin^2 \phi} + \sqrt{r_{pe}^2 - R^2 \sin^2 \phi} \quad (40)$$

After some manipulations, $\phi$ can be derived as

$$\phi = \sin^{-1} \sqrt{\left(\frac{r_{pe}}{R}\right)^2 - \left(\frac{r_{pe}^2 - r_{me}^2 - (vT_{pp})^2}{2RvT_{pp}}\right)^2} \quad (41)$$

Using (2), (31), and (38), the PP probability can be written as

$$P_{\text{PP}} = \begin{cases} 0, & \text{if } \sqrt{r_{pe}^2 - r_{me}^2} > vT_{pp} \\ P\left(2\sqrt{R^2 - r_{me}^2} < d(\alpha) < 2R\right), & \text{if } (r_{me} + r_{pe}) < vT_{pp} \\ P\left(2\sqrt{R^2 - r_{me}^2} < d(\alpha) < 2R\cos\phi\right), & \text{otherwise} \end{cases}$$
$$(42)$$

Complying with the definition of PP rate defined by 3GPP, the PP probability can be redefined as

$$P_{\text{PP,3GPP}} = \frac{P_{\text{PP}}}{1 - P_{\text{NHO}} - P_{\text{HF,m}}} \quad (43)$$

*6) An Extension for Zero Ping-Pong rate*

One notable point is that most PP occurrences are due to the small radius of the pico cell coverage circle because we assumed an ideal HO model with an exact measurement and without fading effect. Therefore, if we keep fast moving users out of small cells as gray-listing solution [48], [49], considering the radius of the pico cell coverage circle, the most PP occurrences can be prevented.

With an extension of keeping fast moving users out of small cells, we can define a special case where the traveling distance during a specific time unit, e.g., $T_{pp}$, is greater than a threshold radius of a circle, i.e., $r_{thresh}$. For the special case, the NHO probability in (32) is

$$P_{NHO} = 1, \quad \text{if } vT_{pp} > r_{thresh} \quad (44)$$

, and the PP probability in (42) is

$$P_{PP} = 0, \quad \text{if } vT_{pp} > r_{thresh} \quad (45)$$

For example, the radius of pico cell coverage circle can be used as the value of $r_{thresh}$.

## IV. RESULTS OF HANDOVER PERFORMANCE ANALYSIS

In this section, we present a methodology to calculate the radii of circles in the HO models discussed in Section III. We derive the radii of circles with different macro-pico distances and differences between DL received signal strengths (RSSs) from macro cell and pico cell. Next, we show the numerical results of HO performance metrics such as HOF and PP probabilities of LTE and ZEUS HO. And we present an extension of keeping fast moving users out of small cells, considering the radius of the coverage of those cells. Then, we discuss the significant findings drawn from the results.

### A. The Calculation of Radii of Circles

Although HO trigger locations and HOF locations are not perfectly circular, it is reasonable to model them geometrically as concentric circles as stated in Section III. We calculated the radii of circles in the HO models following the conventional DL RSS-based cell selection procedure which associates a UE with the cell that provides the strongest DL RSS. The radius of each circle is approximated to the radius of boundary which has a specific value of difference between DL RSSs from macro and pico cell. Therefore, the pico cell coverage circle can be defined as the equal DL RSS boundary, i.e., the boundary along which the DL RSSs from the MNB and the PNB are equal, without pico cell range expansion [50], [51]. RSS and pathloss model in 3GPP LTE HetNet simulation assumptions [6], listed in Table III, is used to calculate the radius of boundary which has a specific value of difference between DL RSSs. We consider three cases of macro-pico distance that are 250 m, 125 m, and 75 m, and the shorter the macro-pico distance, the smaller the radius of pico cell coverage circle yields.

TABLE III
LTE HetNet SIMULATION PARAMETERS

| Parameter | Value |
|---|---|
| Macro BS Tx Power | 46 dBm |
| Macro BS Antenna Gain | 14 dBi |
| Macro Path Loss | 128.1+37.6log10(R[km]) |
| Pico BS Tx Power | 30 dBm |
| Pico BS Antenna Gain | 5 dBi |
| Pico Path Loss | 140.7+36.7log10(R[km]) |
| Carrier Frequency / Bandwidth | 2.0 GHz / 10 MHz |
| Macro-Pico Distance | 75 m, 125 m, 250 m |
| Minimum Path Loss | 35 dB |

TABLE IV
THE RADII OF CIRCLES WITH MACRO-PICO DISTANCES AND $RSS_{DIFF}$

| $RSS_{diff}$ | ISD = 250 m | | | ISD = 125 m | | | ISD = 75 m | | |
|---|---|---|---|---|---|---|---|---|---|
| | m2p | p2m | radius | m2p | p2m | radius | m2p | p2m | radius |
| -8 | 217.32 | 294.64 | 38.66 | 108.90 | 146.88 | 18.99 | 65.44 | 87.93 | 11.25 |
| -6 | 220.71 | 288.54 | 33.92 | 110.57 | 143.90 | 16.66 | 66.44 | 86.17 | 9.87 |
| -4 | 223.80 | 283.37 | 29.79 | 112.10 | 141.37 | 14.64 | 67.34 | 84.68 | 8.67 |
| -3 | 225.23 | 281.08 | 27.93 | 112.80 | 140.25 | 13.72 | 67.76 | 84.02 | 8.13 |
| -2 | 226.59 | 278.97 | 26.19 | 113.47 | 139.21 | 12.87 | 68.16 | 83.41 | 7.62 |
| -1 | 227.88 | 277.01 | 24.56 | 114.11 | 138.25 | 12.07 | 68.54 | 82.84 | 7.15 |
| 0 | 229.11 | 275.20 | 23.04 | 114.72 | 137.37 | 11.32 | 68.90 | 82.32 | 6.71 |
| 1 | 230.28 | 273.52 | 21.62 | 115.29 | 136.54 | 10.62 | 69.24 | 81.83 | 6.29 |
| 2 | 231.39 | 271.96 | 20.28 | 115.84 | 135.78 | 9.97 | 69.57 | 81.38 | 5.91 |
| 3 | 232.44 | 270.51 | 19.04 | 116.36 | 135.07 | 9.36 | 69.87 | 80.96 | 5.54 |
| 4 | 233.44 | 269.17 | 17.86 | 116.85 | 134.41 | 8.78 | 70.17 | 80.57 | 5.20 |
| 6 | 235.27 | 266.75 | 15.74 | 117.75 | 133.22 | 7.74 | 70.70 | 79.87 | 4.58 |
| 8 | 236.92 | 264.66 | 13.87 | 118.56 | 132.20 | 6.82 | 71.18 | 79.26 | 4.04 |

Table IV shows the calculated radii of boundaries which have a specific value of difference between DL RSSs, i.e., $RSS_{diff}$. $RSS_{diff}$ of –8 dB means that RSS from macro cell is better than that from pico cell by 8 dB. The $RSS_{diff}$ of 0 dB means the equal DL RSS. In an inbound HO process, i.e., macro to pico HO, the $RSS_{diff}$ of 8 dB is used as $Q_{out}$ [52], 6 dB as $Q_{in}$ [52], 2 dB or 1 dB as HOM of LTE and HOP event of ZEUS, and 3 dB or 2 dB as HOE event of ZEUS. In an outbound HO process, i.e., pico to macro HO, the $RSS_{diff}$ of –8 dB is used as $Q_{out}$, –6 dB as $Q_{in}$, –2 dB or –1 dB as HOM of LTE and HOP event of ZEUS, and –3 dB or –2 dB as HOE event of ZEUS. We used –4 dB, instead of –6 dB, as $Q_{in}$ in an outbound HO process to derive the numerical results of HO performance because the outbound HO suffers from interferences from neighbor cells more than the inbound HO.

There are two boundaries which has a specific value of difference between DL RSSs, in each direction from the location of the PNB. In Table IV, 'm2p' is the direction getting near to the location of the MNB, and 'p2m' is the opposite direction. For example, in macro-pico distance of 250 m case, the distances from the location of the MNB of two boundaries which have the $RSS_{diff}$ of 0 dB are 229.11 m in 'm2p' direction and 275.20 m in 'p2m' direction. Therefore, we use a half of a distance between two boundaries as the radius of a specific circle. For example, the radius of the pico cell coverage circle is 23.04 m in macro-pico distance of 250 m case, 11.32 m in 125 m case, and 6.71 m in 75 m case.

### B. Example of Handover Scenarios of LTE vs. ZEUS

Fig. 6 shows the profile of DL RSSs from macro cell and pico cell, calculated in the previous sub section, and illustrates example of HO scenarios of LTE and ZEUS. In an inbound HO process, the boundary of HO trigger is assumed as the $RSS_{diff}$ of 2 dB, and that of PDCCH outage as the $RSS_{diff}$ of 6 dB. The size of HO region can be defined as the distance from the location of HO trigger and the location of PDCCH outage. If a UE failed to receive HO CMD before the PDCCH outage, the HOF occurs.

The shorter the macro-pico distance, the smaller the size of HO region yields. Here (X Km/h, Y ms) denotes the UE

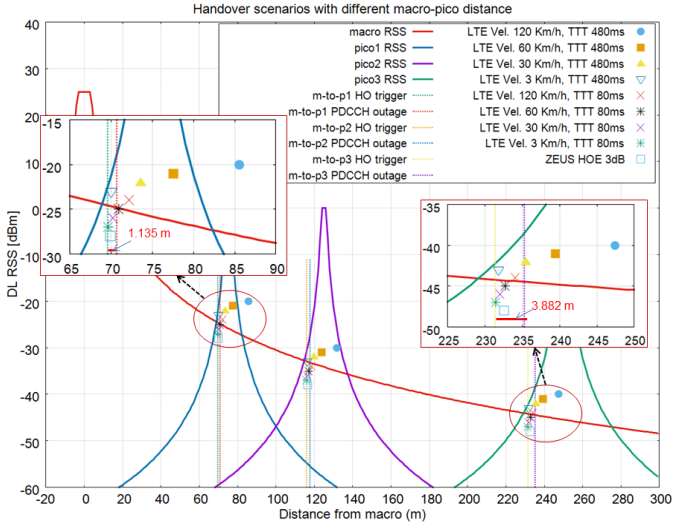

Fig. 6. Example of handover scenarios of LTE vs. ZEUS with different macro-pico distances.

velocity of X Km/h and TTT of Y ms in LTE case. In macro-pico distance of 250 m case, the size of HO region is 3.882 m. The HOF occurs in the cases of (120 Km/h, 480 ms), (60 Km/h, 480 ms), and (30 Km/h, 480 ms), as shown in the right red box. In macro-pico distance of 125 m case, the size of HO region is 1.912 m. In macro-pico distance of 75 m case, the size of HO region is only 1.135 m. The HOF occurs in most cases, except (3 Km/h, 480 ms), (30 Km/h, 80 ms), and (3 Km/h, 80 ms), as shown in the left red box. On the contrary, no HOF occurs with the ZEUS HO regardless of the macro-pico distance and the UE velocity.

### C. The Numerical Results of Handover Performance Metrics

The numerical results of NHO, MUE HOF, PUE HOF, and PP probabilities in three macro-pico distance cases are plotted as a function of UE velocity, in Fig. 7 through 9. The numerical results in macro-pico distance of 250 m case are shown in Fig. 7, 75 m in Fig. 8, and 125 m in Fig. 9. We compare the results in Fig. 7 and Fig. 8 in order to investigate the impact of the UE velocity, macro-pico distance, HOM, and TTT on handover performance. To avoid mutually overlapping of lines, in Fig. 7 and Fig. 8, we omitted the lines of the TTT of 160 ms case. Here (X dB, Y ms) denotes HOM of X dB and TTT of Y ms in LTE case. With an extension for zero PP rate, the radius of pico cell coverage circle in High-speed-ext($R$) case or the radius of $r_{mp}$ circle with HOM of 2 dB in High-speed-ext($r_{mp}$) case is used as the value of $r_{thresh}$.

In short, with LTE HO, the higher the UE velocity or the smaller the macro-pico distance, the higher the HOF rate and PP rate yields, but this is not the case of ZEUS where the HOF rate is always zero regardless of the UE velocity and the macro-pico distance.

*1) The impact of the UE velocity*

With LTE HO, the higher the UE velocity, the higher the HOF rate and PP rate yields. For an example, in macro-pico distance of 250 m and (2 dB, 480 ms) case, the MUE HOF rate is 17 % at 35 Km/h and is increased to 80 % at 120 Km/h. The PUE HOF rate is 52 % at 30 Km/h and is increased to 100 % over 40 Km/h. The PP rate is 3 % at 85 Km/h and is increased to 94 % in 120 Km/h. In all cases, the HOF rate and PP rate are increased proportional to the UE velocity.

On the contrary, with ZEUS HO, the MUE HOF and PUE HOF rate are always zero regardless of the UE velocity. The PP rate is 2 % in 85 Km/h, is increased to 21 % in 120 Km/h, but is lower than that of LTE case in all cases. With an extension for zero PP rate, the PP rate is always zero regardless of the UE velocity, except at 75 Km/h and 80 Km/h case, where the PP rate is under 1 %, plotted by 'ZEUS+High-speed-ext($R$)' line. The PP probabilities of all of three macro-pico distance cases are zero regardless of the mobility speed of a UE, in ZEUS with High-speed-ext($r_{mp}$) case.

*2) The impact of the macro-pico distance*

With LTE HO, the smaller the macro-pico distance, the higher the HOF rate and PP rate yields. For an example, in the UE velocity of 30 Km/h and (2 dB, 480 ms) case, the MUE HOF rate is 0 % in 250 m case and is increased to 77 % in 75 m case. The PUE HOF rate is 52 % in 250 m case and is increased to 100 % in 75 m case. The PP rate is 0 % in 250 m case and is increased to 38 % in 75 m case. In all cases, the HOF rate and PP rate are increased inversely proportional to the macro-pico distance.

On the contrary, with ZEUS HO, the MUE HOF and PUE HOF rate are always zero regardless of the macro-pico distance. The PP rate at 30 Km/h is 0 % in 250 m case, increased to 10 % in 75 m case, but is lower than that of LTE case in all cases. With an extension for zero PP rate, the PP rate is always zero regardless of the macro-pico distance, except at 75 Km/h and 80 Km/h case in 250 m case, as stated earlier.

*3) The impact of the HOM*

With LTE HO, the bigger the HOM, the higher the HOF rate yields. On the contrary, the bigger the HOM, the lower the PP rate yields. For an example, in macro-pico distance of 250 m and in the UE velocity of 60 Km/h, the MUE HOF rate is 65 % in (2 dB, 480 ms) case and is decreased to 48 % in (1 dB, 480 ms) case. The PUE HOF rate is 100 % in (2 dB, 480 ms) case and is decreased to 85 % in (1 dB, 480 ms) case. The PP rate at 90 Km/h is 11 % in (2 dB, 480 ms) case and increased to 34 % in (1 dB, 480 ms) case. However, the exception is observed where the PP rate in (2 dB, 160 ms) case is higher than in (1 dB, 160 ms) case at higher speed. For an example, in macro-pico distance of 125 m case, the PP rate in (2 dB, 160 ms) case is higher than in (1 dB, 160 ms) case above 55 Km/h. That is because a ToS at the pico cell is roughly equal to $T_{pp}$ time unit and an earlier HO into the pico cell with a smaller HOM can increase the ToS. In all other cases, the HOF rate are increased proportional to the HOM and the PP rate are increased inversely proportional to the HOM.

On the contrary, with ZEUS HO, the MUE HOF and PUE HOF rate are always zero regardless of the HOM. The PP rate is lower with bigger HOM for ZEUS HOP event as in LTE case, but it is lower than that of LTE case in all cases.

*4) The impact of the TTT*

With LTE HO, the longer the TTT, the higher the HOF rate yields. On the contrary, the longer the TTT, the lower the PP rate yields. For an example, in macro-pico distance of 250 m and in the UE velocity of 60 Km/h, the MUE HOF rate is 65 % in (2 dB, 480 ms) case and is zero in (2 dB, 80 ms) case. The PUE HOF rate is 100 % in (2 dB, 480 ms) case and is zero in (2 dB, 80 ms) case. The PP rate at 85 Km/h is 3 % in (2 dB, 480 ms) case and is increased to 7 % in (2 dB, 80 ms) case. However, the exception is observed where the PP rate in (1 or 2 dB, 480 ms) case is higher than in (1 or 2 dB, 80 ms) case at higher speed. For an example, in macro-pico distance of 250 m case, the PP rate in (2 dB, 480 ms) case is higher than in (2 dB, 80 ms) case above 90 Km/h. Also, that is because a ToS at the pico cell is roughly equal to $T_{pp}$ time unit and an earlier HO into the pico cell with a shorter TTT can increase the ToS. In all other cases, the HOF rate are increased proportional to the TTT and the PP rate are increased inversely proportional to the TTT.

On the contrary, with ZEUS HO, the TTT is not used basically as stated in Section II. Instead, the TTE is automatically well scaled depending on the UE velocity and the macro-pico distance as shown in Fig. 6.

*D. The significant findings drawn from the results*

The numerical results demonstrate that the ZEUS HO can solve the trade-off between the HOF rate and the PP rate, achieving zero HOF rate without increasing the PP rate. With the ZEUS HO, the probability of MUE HOF and PUE HOF is always zero regardless of the macro-pico distance and the UE velocity. The UE speed has no significant impact on the HO performance, and it is different from the common observation that high-speed UEs suffer much higher HOF rate than low-speed UEs as overall observations in [6]. And, the probability of PP is the lowest in all cases with HOP event of 2 dB and HOE event of 3 dB. It is also different from the common understanding that there is a trade-off between an aggressive HO parameter use to decrease the HOF rate and the amount of PPs as overall observations in [6]. The probability of EHOP, the cost of the ZEUS HO, is marginal and only 6.6% in (ZEUS HOP 2 dB and HOE 3 dB) and 9.0% in (ZEUS HOP 1 dB and HOE 2 dB).

Furthermore, the results show that the ZEUS HO can accomplish zero HOF rate and zero PP rate simultaneously with an extension of keeping fast moving users out of small cells, considering the radius of the coverage of those cells. With the ZEUS HO, we may design a flawless HO algorithm where HOF or PP occurrences are nonexistent.

*E. Future works*

In this article, an abstracted HO model is used in the theoretical analysis, without fading effect and interferences from neighbor cells. For future work, we plan to perform the analysis using a little more complicated HO model, additionally considering the fading effect and interferences. However, the shadowing effect and fast fading are reduced by L3 filtering [46]. Even if the UE speed is 350 km/h, the fluctuation of RSRP can be reduced and is feeble enough to ensure the accuracy of HO decision, with L3 filtering factor $a$ [46] of 1/4, i.e., L3 filter coefficient $k$ [46] of 8, in rural scenario and factor $a$ of 1/8, i.e., coefficient $k$ of 13, in mountainous scenario [53]. But, with the decrease of factor $a$, it can delay the HO decision and increase the HOF rate. Therefore, we need a more study on the balance of reducing the fluctuation of RSRP and keeping the instantaneity of measured RSRP. Also, we need to verify the performance of the ZEUS HO through an extensive simulation of various scenarios that simulates real network deployments. The preliminary simulation results using OPNET LTE Model [54] that compare the HO performance of LTE and ZEUS can be found in [42].

## V. Conclusion

Network densification is regarded as the dominant driver for wireless evolution into the era of 5G. Through the use of a large number of small cells, it boosts the wireless system capacity by providing cell-splitting gains. However, it has increased the complexity of mobility management, and operators have been facing the technical challenges in HO parameter optimization. The trade-off between the HOF rate and the PP rate has further complicated the challenges. We have proposed the ZEUS HO to solve the trade-off, splitting an HO event into an HO preparation event and HO execution event. While the LTE HO is fully network-controlled, the ZEUS HO is hybrid-controlled in that it transfers part of the control of the cell selection at an HO to a UE. The ZEUS HO consists of the network-controlled HO preparation and the UE-controlled HO execution.

We have introduced geometry-based HO models for LTE and ZEUS, considering the HOM, which are used for the HO performance analysis. We have derived the probabilities of NHO, MUE HOF, PUE HOF and PP in closed-form expressions from the analysis, and presented the numerical results of those in three macro-pico distance cases as a function of UE velocity. The numerical results have demonstrated that the ZEUS HO can solve the trade-off between the HOF rate and the PP rate, achieving zero HOF rate without increasing the PP rate. With the ZEUS HO, the probability of HOF is always zero regardless of the macro-pico distance and the UE velocity, only with the marginal cost of the extra HO preparation. Furthermore, we have showed that the ZEUS can achieve zero HOF and zero PP rate simultaneously with an extension of keeping fast moving users out of small cells, considering the radius of the coverage of those cells. With the ZEUS HO, we may design a flawless HO algorithm where HOF or PP occurrences are nonexistent. The ZEUS HO can relieve the operators' efforts for network and HO parameter tuning. The ZEUS HO is the most reliable and a promising HO control algorithm in 5G networks.

Acknowledgment

The authors specially thank Karthik Vasudeva, İsmail Güvenç, and David López Pérez, for their previous works and valuable discussions on the theoretical analysis of the HO performance.

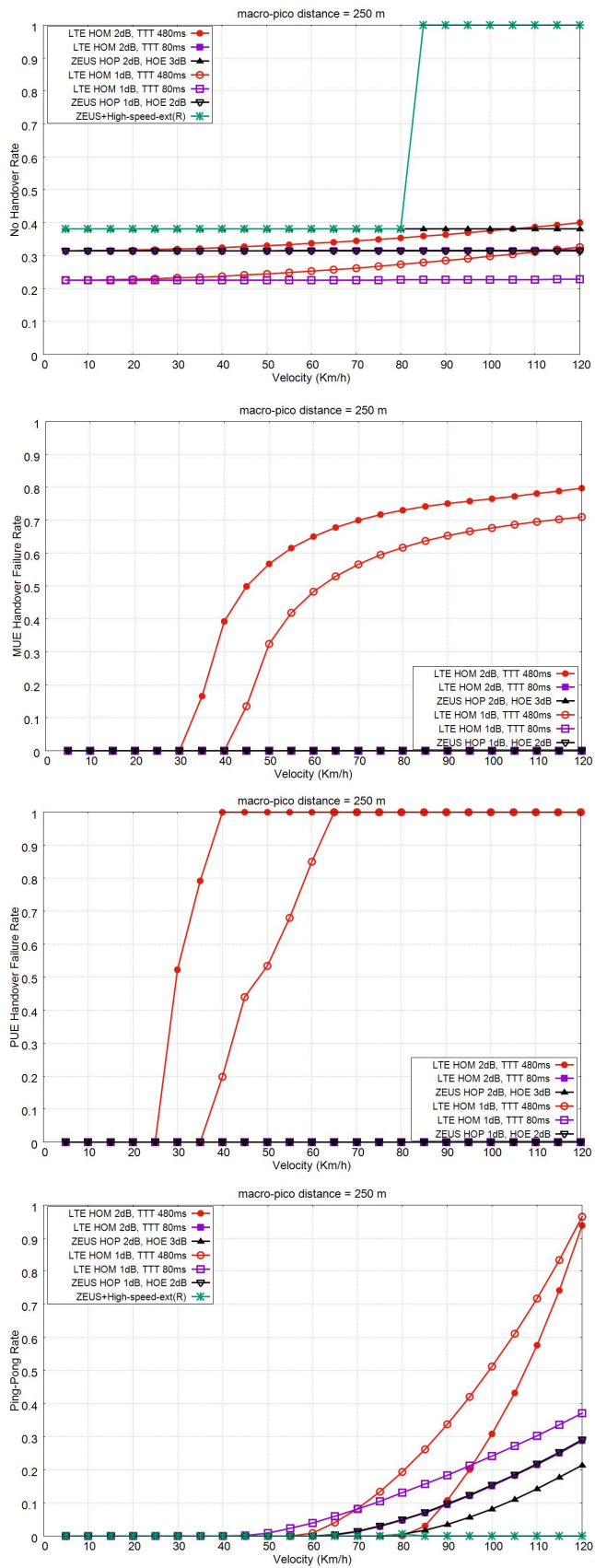

Fig. 7. Results for NHO, MUE HOF, PUE HOF, and PP rates (macro-pico distance = 250 m).

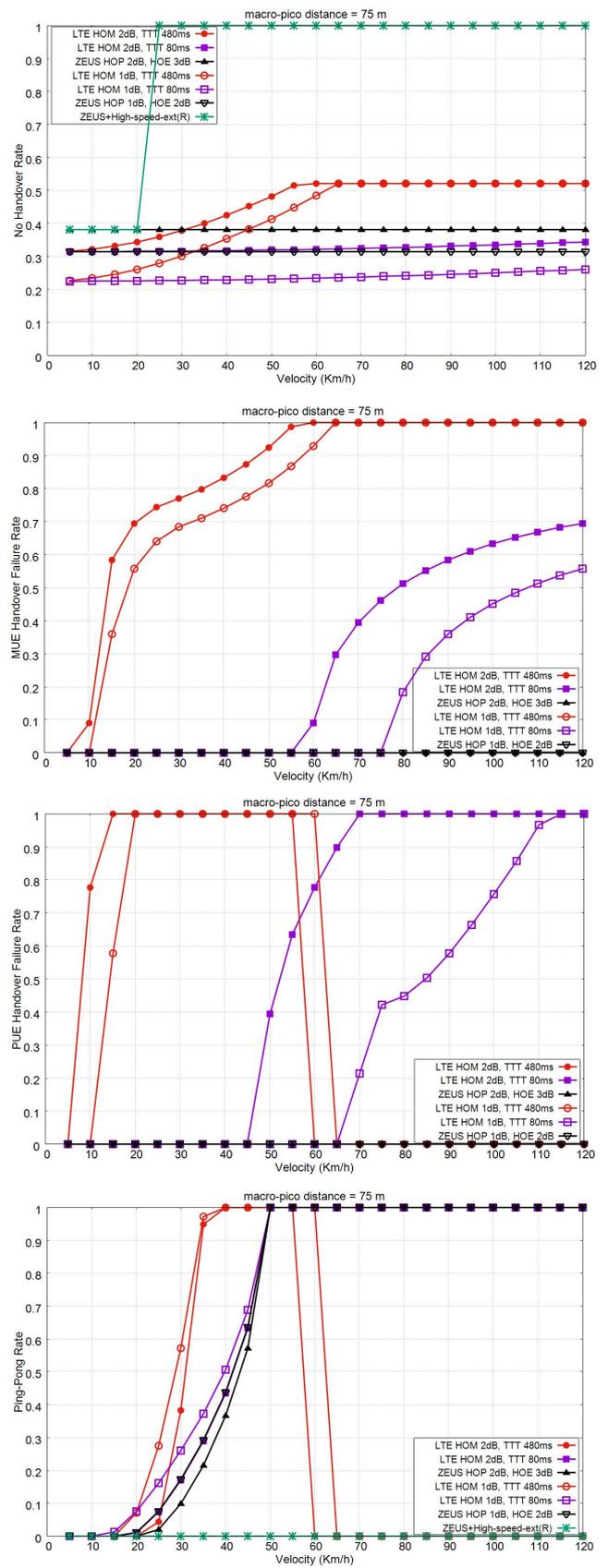

Fig. 8. Results for NHO, MUE HOF, PUE HOF, and PP rates (macro-pico distance = 75 m).

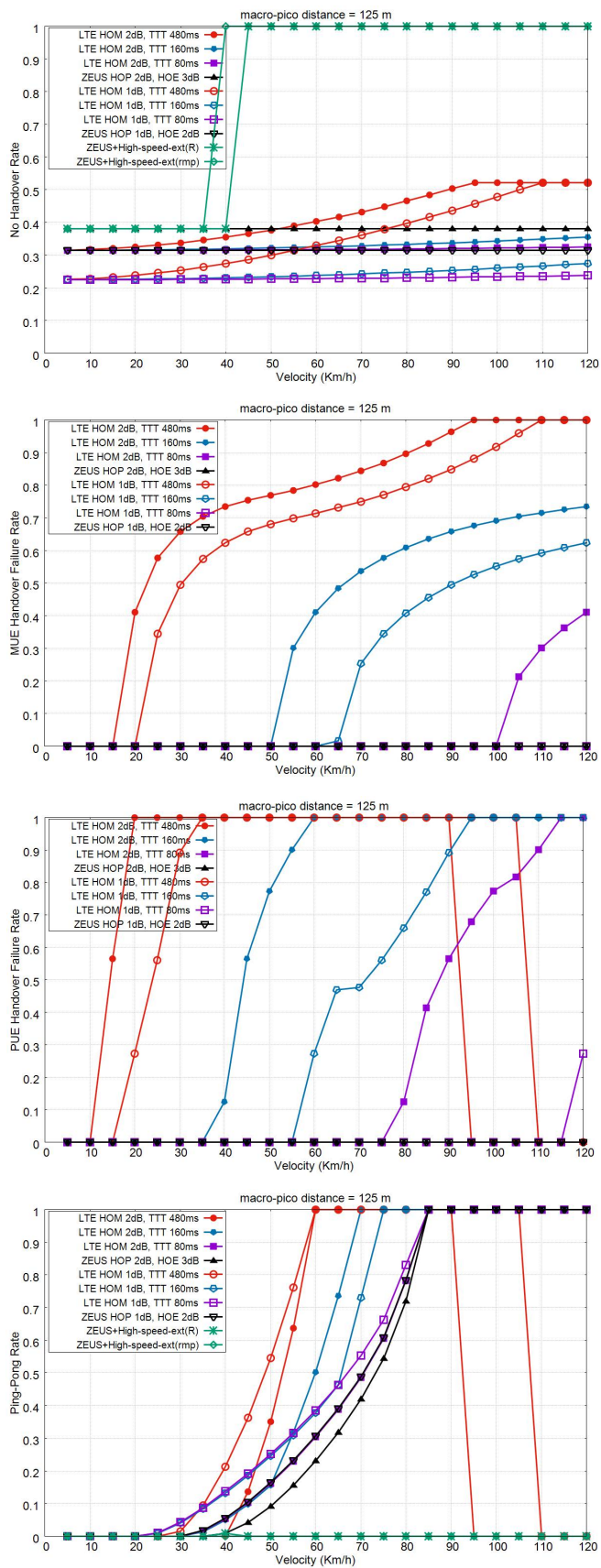

Fig. 9. Results for NHO, MUE HOF, PUE HOF, and PP rates (macro-pico distance = 125 m).

REFERENCES

[1] N. Bhushan, J. Li, D. Malladi, R. Gilmore, D. Brenner, A. Damnjanovic, R. T. Sukhavasi, and S. Geirhofer, "Network Densification: The Dominant Theme for Wireless Evolution into 5G," *IEEE Communications Magazine,* vol. 52, no. 2, Feb. 2014, pp. 82–89.
[2] *Future technology trends of terrestrial IMT systems*, Report ITU-R M.2320-0, Nov. 2014.
[3] D. López-Pérez, M. Ding, H. Claussen, and A. H. Jafari, "Towards 1 Gbps/UE in Cellular Systems: Understanding Ultra-Dense Small Cell Deployments," *IEEE Communications Surveys & Tutorials*, vol. 17, no. 4, Nov. 2015, pp. 2078–2101.
[4] D. López-Pérez, I. Guvenc, and X. Chu, "Mobility Management Challenges in 3GPP Heterogeneous Networks," *IEEE Communications Magazine,* vol. 50, no. 12, Dec. 2012, pp. 70–78.
[5] X. Zheng, J. Yu, Z. Wei, H. Hu, Y. Yang, and H. Chen, "Mobility Management and Performance Optimization in Next Generation Heterogeneous Mobile Networks", in *Heterogeneous Cellular Networks*, John Wiley & Sons, 2013, pp. 165–197.
[6] 3GPP TR 36.839, *Evolved Universal Terrestrial Radio Access (E-UTRA); Mobility enhancements in heterogeneous networks,* 2012.
[7] 3GPP TR 36.842, *Study on Small Cell enhancements for E-UTRA and E-UTRAN; Higher layer aspects,* 2013.
[8] J. Moon, J. Jung, S. Lee, A. Nigam, and S. Ryoo, "On the Trade-Off between Handover Failure and Small Cell Utilization in Heterogeneous Networks," *IEEE Int. Workshop on Advanced PHY and MAC Techniques for Super Dense Wireless Networks (co-located with IEEE ICC),* London, UK, 2015, pp. 10348–10353.
[9] J. Acharya, L. Gao, and S. Gaur, "Dense Small Cell Deployments", in *Heterogeneous Networks in LTE-Advanced*, John Wiley & Sons, 2014, pp. 205–229.
[10] 3GPP RAN2 R2-140089, *Mobility performance in real network*, Qualcomm Incorporated, RAN2#85, Feb. 2014.
[11] H. Chen, S. Jin, H. Hu, Y. Yang, D. López-Pérez, I. Guvenc, and X. Chu, "Mobility and handover management", in *Heterogeneous Cellular Networks: Theory, Simulation and Deployment*, Cambridge University Press, 2013, pp. 245–283.
[12] 3GPP RAN2 R2-131667, *Shorter T310 at handover failure*, Ericsson, RAN2#82, May 2013.
[13] 3GPP RAN RP-142036, *New SI proposal: Study on Mobility Enhancements for LTE,* Ericsson, RAN#66, Dec. 2014.
[14] Y. Zhou, Z. Lei, and S. Wong, "Evaluation of Mobility Performance in 3GPP Heterogeneous Networks," in *Proc. IEEE 79th VTC*, Seoul, Korea, 2014, pp. 1–5.
[15] D. López-Pérez, I. Guvenc, and X. Chu, "Theoretical Analysis of Handover Failure and Ping-Pong Rates for Heterogeneous Networks," *IEEE Int. Workshop on Small Cell Wireless Networks (co-located with IEEE ICC)*, Ottawa, Canada, 2012, pp. 6774–6779.
[16] A. Jensen, K. Pedersen, M. Lauridsen, P. Mogensen, and J. Faaborg, "LTE HetNet Mobility Performance Through Emulation with Commercial Smartphones," *in Proc. IEEE 79th VTC*, Seoul, Korea, 2014, pp. 1–5.
[17] J. Zhang, J. Feng, C. Liu, X. Hong, X. Zhang, and W. Wang, "Mobility Enhancement and Performance Evaluation for 5G Ultra Dense Networks," *in Proc. IEEE WCNC*, New Orleans, USA, 2015, pp. 1793–1798.
[18] S. Lee, J. Jung, J. Moon, A. Nigam, and S. Ryoo, "Mobility Enhancement of Dense Small-Cell Network," *in Proc. 12th IEEE CCNC*, Las Vegas, USA, 2015, pp. 297–303.
[19] 3GPP RAN RP-122007, *New WI proposal: Hetnet Mobility Enhancements for LTE,* Alcatel-Lucent, RAN#58, Dec. 2012.
[20] 3GPP RAN2 R2-132990, *Report of email discussion [82#16] [LTE/Het-Net] Mobility Robustness,* Alcatel-Lucent, RAN2#83, Aug. 2013.
[21] 3GPP RAN2 R2-133449, *Report of email discussion [83#12] [LTE/Het-Net] Evaluate UE based solutions for mobility robustness,* Alcatel-Lucent, RAN2#83bis, Oct. 2013.
[22] 3GPP RAN RP-140359, *RAN2 agreed CRs on Core part: Hetnet Mobility Enhancements for LTE,* RAN#63, March 2014.
[23] 3GPP RAN RP-141505, *RAN2 agreed CRs on Core part: Hetnet Mobility Enhancements for LTE,* RAN#65, Sept. 2014.
[24] 3GPP RAN RP-122033, *New Study Item Description: Small Cell enhancements for E-UTRA and E-UTRAN – Higher-layer aspects,* NTT DoCoMo, RAN#58, Dec. 2012.
[25] 3GPP RAN2 R2-132682, *Mobility information at RRC Connection Establishment*, Ericsson, RAN2#83, Aug. 2013.

[26] 3GPP RAN2 R2-130961, *Evaluation on adjusting parameters for the handover type,* CATT, RAN2#81bis, April 2013.
[27] 3GPP RAN2 R2-123431, Context Fetch for RRC Connection Re-establishment in HetNets, Alcatel-Lucent, RAN2#79, Aug. 2012.
[28] Q. Kuang, J. Belschner, M. Lossow, P. Arnold, and O. Ramos-Cantor, "Mobility Performance of LTE-Advanced Heterogeneous Networks with Control Channel Protection," in *Proc. WTC,* Berlin, Germany, 2014, pp. 1–6.
[29] K. Adachi, J. Joung, Y. Zhou, and S. Sun, "A Distributed Resource Reservation Scheme for Handover Failure Reduction," *IEEE Wireless Communications Letters,* vol. 4, no. 5, Oct. 2015, pp. 537–540.
[30] X. Zhang, Z. Xiao, S. Mahato, E. Liu, B. Allen, and C. Maple, "Dynamic user equipment-based hysteresis-adjusting algorithm in LTE femtocell networks," *IET Communications,* vol. 8, no. 17, 2014, pp. 3050–3060.
[31] C. Rosa, K. Pedersen, H. Wang, P. Michaelsen, S. Barbera, E. Malkamaki, T. Henttonen, and B. Sebire, "Dual Connectivity for LTE Small Cell Evolution: Functionality and Performance Aspects," *IEEE Communications Magazine,* vol. 54, no. 6, June 2016, pp. 137–143.
[32] 3GPP RAN2 R2-134171, *Effect of Handover Delay on Handover Failure and Ping-Pong in Dense HetNet,* Samsung, RAN2#84, Nov. 2013.
[33] 3GPP RAN RP-142047, *New WID: Mobility Enhancements for LTE,* Mediatek Inc., RAN#66, Dec. 2014.
[34] 3GPP RAN RP-142005, *New SI: Partial control of Cell Management for UEs in LTE,* Nokia Networks, RAN#66, Dec. 2014.
[35] K. I. Pedersen, P. H. Michaelsen, C. Rosa, and S. Barbera, "Mobility Enhancements for LTE-Advanced Multilayer Networks with Inter-Site Carrier Aggregation," *IEEE Communications Magazine,* vol. 51, no. 5, May 2013, pp. 64–71.
[36] *METIS Deliverable D4.3: Final Report on Network-Level Solutions*, ICT-317669-METIS/D4.3, Jan. 2015.
[37] K. Vasudeva, M. Simsek, and I. Guvenc, "Analysis of Handover Failures in HetNets with Layer-3 Filtering," in *Proc. WCNC,* Istanbul, Turkey, 2014, pp. 2647–2652.
[38] K. Vasudeva, M. Simsek, D. López-Pérez, and I. Guvenc, "Impact of Channel Fading on Mobility Management in Heterogeneous Networks," *IEEE Int. Workshop on Advanced PHY and MAC Techniques for Super Dense Wireless Networks (co-located with IEEE ICC),* London, UK, 2015, pp. 10272–10277.
[39] H. Park, A. Park, J. Lee, and B. Kim, "Two-Step Handover for LTE HetNet mobility enhancements," in *Proc. ICTC,* Jeju, Korea, 2013, pp. 763–766.
[40] 3GPP RAN2 R2-134432, *Early HO CMD with Ping-Pong Avoidance, further information,* ETRI, RAN2#84, Nov. 2013.
[41] H. Park, and Y. Choi, "Taking advantage of multiple handover preparations to improve handover performance in LTE networks," in *Proc. FGCN,* Hainan, China, 2014, pp. 9–12.
[42] H. Park, Y. Choi, B. Kim, and J. Lee, "LTE Mobility Enhancements for Evolution into 5G," *ETRI Journal*, vol. 37, no. 6, Dec. 2015, pp. 1065–1076.
[43] 3GPP RAN2 R2-131195, *Main Directions for Intra-frequency HetNet mobility enhancements,* Samsung, RAN2#81bis, April 2013.
[44] 3GPP TS 36.300, *Evolved Universal Terrestrial Radio Access (E-UTRA) and Evolved Universal Terrestrial Radio Access Network (E-UTRAN); Overall description; Stage 2,* 2015.
[45] S. Yi, S. Chun, Y. Lee, S. Park, and S. Jung, "Radio Resource Control (RRC)", in *Radio Protocols for LTE and LTE-Advanced,* JohnWiley & Sons, 2012, pp. 47–85.
[46] 3GPP TS 36.331, *Evolved Universal Terrestrial Radio Access (E-UTRA); Radio Resource Control (RRC),* 2015.
[47] A. M. Mathai, "Random Points and Random Distances", in *An Introduction to Geometrical Probability: Distributional Aspects with Applications*, Gordon and Breach Science Publishers, 1999, pp. 171–186.
[48] 3GPP RAN2 R2-130550, HetNet co-channel optimization based on Gray-listing and eMSE, Nokia Siemens Networks, RAN2#81, Feb. 2013.
[49] S. Barbera, P. H. Michaelsen, M. Saily, K. Pedersen, "Improved mobility performance in LTE co-channel hetnets through speed differentiated enhancements," *IEEE Workshop on Heterogeneous, Multi-Hop, Wireless and Mobile Networks (co-located with IEEE GLOBECOM),* Anaheim, USA, Dec. 2012, pp. 426–430.
[50] D. López-Pérez, X. Chu, and I. Guvenc, "On the Expanded Region of Picocells in Heterogeneous Networks," *IEEE Journal of Selected Topics in Signal Processing*, vol. 6, no. 3, June 2012, pp. 281–294.
[51] 3GPP RAN2 R2-114950, *Discussion on the mobility performance enhancement for co-channel HetNet deployment,* ZTE, RAN2#75bis, Oct. 2011.
[52] 3GPP TS 36.133, *Evolved Universal Terrestrial Radio Access (E-UTRA); Requirements for support of radio resource management,* 2015.
[53] A. Zhang, J. Miao, Y. Zhang, Y. Liu, and S. Lu, "Impact Analysis of Layer 3 Filtering on LTE Handover Performance in High-speed Railway Scenario," *International Journal of Digital Content Technology and its Applications(JDCTA),* vol. 7, no. 7, April 2013, pp. 492–499.
[54] OPNET (Riverbed application and network performance management solutions).Available: http://www.riverbed.com/products/performance-management-control/opnet.html